\documentclass[a4paper,english,aps,pra,preprint,groupedaddress]{revtex4}
\usepackage[T2A]{fontenc}
\usepackage[latin9]{inputenc}
\usepackage{color}
\usepackage{babel}

\usepackage{units}
\usepackage{amsmath}
\usepackage{graphicx}
\usepackage{amssymb}
\usepackage[unicode=true, pdfusetitle,
 bookmarks=true,bookmarksnumbered=true,bookmarksopen=false,
 breaklinks=false,pdfborder={0 0 0},backref=false,colorlinks=true]
 {hyperref}
\hypersetup{
 citecolor=blue,pdfstartview=FitH}
 
\makeatletter
\@ifundefined{textcolor}{}
{%
 \definecolor{BLACK}{gray}{0}
 \definecolor{WHITE}{gray}{1}
 \definecolor{RED}{rgb}{1,0,0}
 \definecolor{GREEN}{rgb}{0,1,0}
 \definecolor{BLUE}{rgb}{0,0,1}
 \definecolor{CYAN}{cmyk}{1,0,0,0}
 \definecolor{MAGENTA}{cmyk}{0,1,0,0}
 \definecolor{YELLOW}{cmyk}{0,0,1,0}
 }

\makeatother

\begin{document}

\title{Spin-dependent part of $p\overline{p}$ interaction cross section
and Nijmegen potential.}

\author{V.~F.~Dmitriev}

\email{V.F.Dmitriev@inp.nsk.su}

\author{A.~I.~Milstein}

\email{A.I.Milstein@inp.nsk.su}

\author{S.~G.~Salnikov}

\email{salsergey@gmail.com}

\affiliation{Budker Institute of Nuclear Physics, 630090 Novosibirsk, Russia}

\affiliation{Novosibirsk State University, 630090 Novosibirsk, Russia}

\date{\today}
\begin{abstract}
Low energy $p\bar{p}$~interaction is considered taking into account
the polarization of both particles. The corresponding cross sections
are calculated using the Nijmegen nucleon-antinucleon optical potential.
Then they are applied to the analysis of the polarization buildup
which is due to the interaction of stored antiprotons with polarized
protons of a hydrogen target. It is shown that, at realistic parameters
of a storage ring and a target, the filtering mechanism may provide
a noticeable polarization in a time comparable with the beam lifetime.
\end{abstract}
\maketitle

\section{Introduction}

An extensive research program with polarized antiprotons has been
proposed recently by the PAX Collaboration~\cite{PAX05}. This program
has initiated a discussion of various methods to polarize stored antiprotons.
One of the methods is to use multiple scattering on a polarized hydrogen
target. If all particles remain in the beam (scattering angle is smaller
than acceptance angle $\theta_{\textrm{acc}}$), only spin flip can
lead to polarization buildup, as was shown in Refs.~\cite{MilStr05,NikPav06}.
However, spin-flip cross section is negligibly small in both cases
of proton-\hspace{0cm}antiproton~\cite{MilStr05} and electron-\hspace{0cm}antiproton~\cite{MilSalStr08}
scattering. Hence the most realistic method is spin filtering~\cite{Csonka68}.
This method implements the dependence of scattering cross section
on orientation of particles spins. Therefore number of antiprotons
scattered out of the beam after interaction with a polarized target
depends on their spins, which results in the polarization buildup.
Interaction with atomic electrons can't provide noticeable polarization
because in this case antiprotons will scatter only in small angles
and all antiprotons remain in the beam~\cite{MilStr05}. Thus it
is necessary to study proton-\hspace{0cm}antiproton scattering.

At present, quantum chromodynamics can't give reliable predictions
for $p\overline{p}$~cross section below~$\unit[1]{GeV}$ and different
phenomenological models are usually used for numerical estimations.
As a result, the cross sections obtained are model-\hspace{0cm}dependent.
All models are based on fitting of experimental data for scattering
of unpolarized particles. These models give similar predictions for
spin-\hspace{0cm}independent part of the scattering cross sections,
but predictions for spin-\hspace{0cm}dependent parts may differ drastically.

Different nucleon-\hspace{0cm}antinucleon potentials have similar
behavior at large distance (\mbox{$r\gtrsim\unit[1]{fm}$}) because
long\nobreakdash-range potentials are obtained by applying G\nobreakdash-\hspace{0cm}parity
transformation to well\nobreakdash-known nucleon-\hspace{0cm}nucleon
potential. The most important difference between nucleon-\hspace{0cm}antinucleon
and nucleon-\hspace{0cm}nucleon scattering is existence of annihilation
channels. A phenomenological description of annihilation is usually
based on an optical potential of the form

\begin{equation}
V_{N\overline{N}}=U_{N\overline{N}}-iW_{N\overline{N}}.\end{equation}
Imaginary part of this potential describes annihilation into mesons
and is important at small distance. The process of annihilation has
no uniform description, and short\nobreakdash-range potentials in
various models are different.

Spin-\hspace{0cm}dependent part of the cross section of $p\overline{p}$~interaction
was previously calculated in Ref.~\cite{DmitMilStr07} using the
Paris potential. In~Ref.~\cite{DmitMilStr07} a possibility to obtain
a noticeable beam polarization in a reasonable time was also investigated.
Similar calculations were performed in Ref.~\cite{PAX09} where various
forms of Julich potentials were explored. Note that the contribution
of interference between the Coulomb and strong amplitudes to the scattering
cross section has been omitted in Ref.~\cite{PAX09}. In the present
paper, we calculate the spin-\hspace{0cm}dependent part of the cross
section of $p\overline{p}$~scattering using the Nijmegen model and
analyze the polarization buildup which is due to the interaction of
stored antiprotons with polarized protons.

\section{Cross sections}

It is convenient to calculate the cross section in the center-of-mass
frame, where antiproton and proton have momenta $\boldsymbol{p}$
and $-\boldsymbol{p}$, respectively. In the nonrelativistic approximation
($p\ll M$, where $M$ is the nucleon mass), the antiproton momentum
in the lab frame is $\boldsymbol{p}_{\mathrm{lab}}=2\boldsymbol{p}$.
Therefore the acceptance angle (maximum scattering angle when antiprotons
remain in the beam) in the lab frame is connected with the acceptance
angle in the center-of-mass frame by the relation $\theta_{\textrm{acc}}=2\theta_{\textrm{acc}}^{(l)}$.

The $p\overline{p}$~scattering process has several channels: elastic
scattering (\mbox{$p\overline{p}\to p\overline{p}$}), charge exchange
(\mbox{$p\overline{p}\to n\overline{n}$}), and annihilation into
mesons (\mbox{$p\overline{p}\to mesons$}). As explained above,
noticeable polarization can be obtained only if some antiprotons are
dropped out of the beam. The corresponding cross section can be written
in the form

\begin{equation}
\sigma=\sigma_{\textrm{el}}+\sigma_{\textrm{cex}}+\sigma_{\textrm{ann}},\end{equation}
where $\sigma_{\textrm{cex}}$~is the charge exchange cross section,
$\sigma_{\textrm{ann}}$~is the annihilation cross section, and $\sigma_{\textrm{el}}$~is
the elastic cross section integrated over scattering angle from $\theta_{\textrm{acc}}$
to~$\pi$. All cross sections are summed up over final spin states.
The cross section $\sigma_{\textrm{el}}$~includes pure Coulomb cross
section, hadronic cross section, and interference term, which can't
be omitted.

Spin-\hspace{0cm}dependent cross section can be written in the form

\begin{equation}
\sigma=\sigma_{0}+(\boldsymbol{\zeta}_{1}\cdot\boldsymbol{\zeta}_{2})\sigma_{1}+(\boldsymbol{\zeta}_{1}\cdot\boldsymbol{v})(\boldsymbol{\zeta}_{2}\cdot\boldsymbol{v})(\sigma_{2}-\sigma_{1}),\label{eq:CrossSection}\end{equation}
where $\boldsymbol{\zeta}_{1}$ and $\boldsymbol{\zeta}_{2}$ are
unit vectors collinear to the particles spins, and $\boldsymbol{v}=\boldsymbol{p}/p$
is unit momentum vector. Here $\sigma_{0}$~is the spin-\hspace{0cm}independent
cross section, $\sigma_{1}$~describes spin effects in the case when
both vectors of polarization are perpendicular to $\boldsymbol{v}$,
and $\sigma_{2}$~describes spin effects in the case when all three
vectors are collinear. We direct quantization axis along the vector
$\boldsymbol{v}$ and express the cross sections~\eqref{eq:CrossSection}
via cross sections $\Sigma_{S\mu}$ calculated for states with total
spin $S$ and projection of total angular momentum $\mu$:

\begin{align}
\sigma_{0} & =\frac{1}{2}\Sigma_{11}+\frac{1}{4}\left(\Sigma_{10}+\Sigma_{00}\right),\nonumber \\
\sigma_{1} & =\frac{1}{4}\left(\Sigma_{10}-\Sigma_{00}\right),\label{eq:SpinCrossSections}\\
\sigma_{2} & =\frac{1}{2}\Sigma_{11}-\frac{1}{4}\left(\Sigma_{10}+\Sigma_{00}\right).\nonumber \end{align}
Here we use the relation $\Sigma_{1-1}=\Sigma_{11}$.

The potential of proton-\hspace{0cm}antiproton interaction is a sum
of the Coulomb potential and optical Nijmegen potential~\cite{TimmRijSw94}.
Therefore the amplitude of elastic scattering can be written as a
sum of the Coulomb amplitude and strong amplitude, which doesn't coincide
with the amplitude calculated in the absence of the Coulomb field.
Strong amplitude is not singular at small scattering angles, so that
we can integrate hadronic cross section over the whole range from
$0$ to $\pi$. However, finite $\theta_{\mathrm{acc}}$ should be
taken into account at calculation of the Coulomb cross section and
the interference between the Coulomb and strong amplitudes. In the
nonrelativistic limit, the Coulomb amplitude is spin-independent and
has the form\begin{equation}
F_{1\mu}^{\mathrm{C}}=F_{00}^{\mathrm{C}}=F^{\mathrm{C}}(\theta)=\frac{\alpha}{4vp\sin^{2}(\theta/2)}\exp\left\{ -2i\eta\ln\left[\sin(\theta/2)\right]+2i\chi_{0}\right\} ,\end{equation}
where $\chi_{L}=\arg\Gamma\left(L+1+i\eta\right)$ are the Coulomb
phases, $\eta=-\frac{\alpha}{v_{\mathrm{lab}}}$ is the Sommerfeld
parameter, $v_{\mathrm{lab}}=p_{\mathrm{lab}}/M$, and $\alpha$ is
the fine structure constant.

For the strong elastic triplet scattering amplitude, we have

\begin{equation}
\begin{array}{l}
{\displaystyle \vphantom{\biggl|}F_{1\mu}^{\mathrm{el}}=\frac{i\sqrt{4\pi}}{2p}\sum_{m,L,J}C_{Lm,1\mu-m}^{J\mu}R_{L\mu}^{J}Y_{Lm}(\theta,\varphi),}\\
{\displaystyle \vphantom{\biggl|}R_{L\mu}^{J}=\sum_{L'}(-1)^{\frac{L-L'}{2}}\sqrt{2L'+1}C_{L'0,1\mu}^{J\mu}\exp\left(i\chi_{L}+i\chi_{L'}\right)\left(\delta_{LL'}-S_{LL'}^{J}\right).}\end{array}\label{eq:StrongElasticTriplet}\end{equation}
The sum over $L,\, L'$ is performed under conditions \mbox{$L,L'=J,J\pm1$}
and \mbox{$\left|L-L'\right|=0,2$}. Strong singlet amplitude reads

\begin{equation}
F_{00}^{\mathrm{el}}=\frac{i\sqrt{4\pi}}{2p}\sum_{L}\sqrt{2L+1}\exp(2i\chi_{L})\left(1-S_{L}\right)Y_{L0}(\theta,\varphi).\label{eq:StrongElasticSinglet}\end{equation}
Here $S_{LL'}^{J}$ and $S_{L}$ are partial elastic triplet and singlet
scattering amplitudes, respectively, $Y_{Lm}(\theta,\varphi)$ are
the spherical functions and $C_{Lm,1\mu-m}^{J\mu}$ are the Clebsch\nobreakdash-Gordan
coefficients.

Charge exchange scattering amplitudes have the form

\begin{equation}
\begin{array}{l}
{\displaystyle \vphantom{\biggl|}F_{1\mu}^{\mathrm{cex}}=-\frac{i\sqrt{4\pi}}{2p}\sum_{m,L,J}C_{Lm,1\mu-m}^{J\mu}\widetilde{R}_{L\mu}^{J}Y_{Lm}(\theta,\varphi),}\\
{\displaystyle \vphantom{\biggl|}\widetilde{R}_{L\mu}^{J}=\sum_{L'}(-1)^{\frac{L-L'}{2}}\sqrt{2L'+1}C_{L'0,1\mu}^{J\mu}\exp\left(i\chi_{L'}\right)\widetilde{S}_{LL'}^{J}}\end{array}\end{equation}
and

\begin{equation}
F_{00}^{\mathrm{cex}}=-\frac{i\sqrt{4\pi}}{2p}\sum_{L}\sqrt{2L+1}\exp(i\chi_{L})\widetilde{S}_{L}Y_{L0}(\theta,\varphi).\end{equation}
Here $\widetilde{S}_{LL'}^{J}$ and $\widetilde{S}_{L}$ are partial
charge exchange triplet and singlet scattering amplitudes, respectively.

The cross sections $\Sigma_{1\mu}$ and $\Sigma_{00}$ can be represented
as a sum of pure Coulomb cross sections $\Sigma_{1\mu}^{\mathrm{C}}$
and $\Sigma_{00}^{\mathrm{C}}$, hadronic contributions $\Sigma_{1\mu}^{\mathrm{h}}$
and $\Sigma_{00}^{\mathrm{h}}$, and interference terms $\Sigma_{1\mu}^{\mathrm{int}}$
and $\Sigma_{00}^{\mathrm{int}}$. For the Coulomb contribution we
have

\begin{equation}
\Sigma_{1\mu}^{\mathrm{C}}=\Sigma_{00}^{\mathrm{C}}=\frac{\pi\alpha^{2}}{(vp\theta_{\mathrm{acc}})^{2}},\label{eq:CrossSectionCoulomb}\end{equation}
where smallness of $\theta_{\mathrm{acc}}$ is taken into account.
The total hadronic cross section can be calculated using the optical
theorem

\global\long\def\re#1{\mathrm{Re}#1}

\begin{equation}
\begin{array}{l}
{\displaystyle \Sigma_{1\mu}^{\mathrm{h}}=\frac{2\pi}{p^{2}}\sum_{L,J}\sqrt{2L+1}C_{L0,1\mu}^{J\mu}\re{R_{L\mu}^{J}},}\\
{\displaystyle {\displaystyle \Sigma_{00}^{\mathrm{h}}=\frac{2\pi}{p^{2}}\sum_{L}\left(2L+1\right)\re{\left[\exp(2i\chi_{L})\left(1-S_{L}\right)\right]}.}}\end{array}\label{eq:CrossSectionHadron}\end{equation}
The interference contributions can be calculated in the logarithmic
approximation,\global\long\def\im#1{\mathrm{Im}#1}
\begin{multline}
\Sigma_{1\mu}^{\mathrm{int}}=-\frac{2\pi\alpha}{vp^{2}}\ln\left(\frac{2}{\theta_{\mathrm{acc}}}\right)\sum_{L,J}\sqrt{2L+1}C_{L0,1\mu}^{J\mu}\times{}\\
\shoveright{\times\left\{ \im{\left[\exp(-2i\chi_{0})R_{L\mu}^{J}\right]}+\frac{\alpha}{2v}\ln\left(\frac{2}{\theta_{\mathrm{acc}}}\right)\re{\left[\exp(-2i\chi_{0})R_{L\mu}^{J}\right]}\right\} ,}\\
\shoveleft{\Sigma_{00}^{\mathrm{int}}=-\frac{2\pi\alpha}{vp^{2}}\ln\left(\frac{2}{\theta_{\mathrm{acc}}}\right)\sum_{L}\left(2L+1\right)\times}\\
{}\times\left\{ \im{\left[\exp\left(2i\left(\chi_{L}-\chi_{0}\right)\right)\left(1-S_{L}\right)\right]}+\frac{\alpha}{2v}\ln\left(\frac{2}{\theta_{\mathrm{acc}}}\right)\re{\left[\exp\left(2i\left(\chi_{L}-\chi_{0}\right)\right)\left(1-S_{L}\right)\right]}\right\} .\label{eq:CrossSectionInterference}\end{multline}
The hadronic contributions to the elastic cross sections have the
form

\begin{equation}
\begin{array}{l}
{\displaystyle \Sigma_{1\mu}^{\mathrm{el}}=\frac{\pi}{p^{2}}\sum_{L,J}\left|R_{L\mu}^{J}\right|^{2},}\\
{\displaystyle {\displaystyle \Sigma_{00}^{\mathrm{el}}=\frac{\pi}{p^{2}}\sum_{L}\left(2L+1\right)\left|1-S_{L}\right|^{2}.}}\end{array}\label{eq:CrossSectionElastic}\end{equation}
The charge exchange cross sections are given by

\begin{equation}
\begin{array}{l}
{\displaystyle \Sigma_{1\mu}^{\mathrm{cex}}=\frac{\pi}{p^{2}}\sum_{L,J}\left|\widetilde{R}_{L\mu}^{J}\right|^{2},}\\
{\displaystyle {\displaystyle \Sigma_{00}^{\mathrm{cex}}=\frac{\pi}{p^{2}}\sum_{L}\left(2L+1\right)\left|\widetilde{S}_{L}\right|^{2}.}}\end{array}\label{eq:CrossSectionChargeExchange}\end{equation}

\section{Numerical results}

We follow the method of calculations of scattering amplitudes $S$
described in Refs.~\cite{TimmRijSw94,NagRijSw78}. To calculate $S$\nobreakdash-\hspace{0cm}matrix
and cross sections~(\ref{eq:CrossSectionCoulomb}~--~\ref{eq:CrossSectionChargeExchange}),
we use formula~(15) of Ref.~\cite{TimmRijSw94}. The partial cross
sections obtained are in agreement with the values from Table~V of
Ref.~\cite{TimmRijSw94}.

\begin{figure}
\begin{centering}
\includegraphics[height=4.7cm]{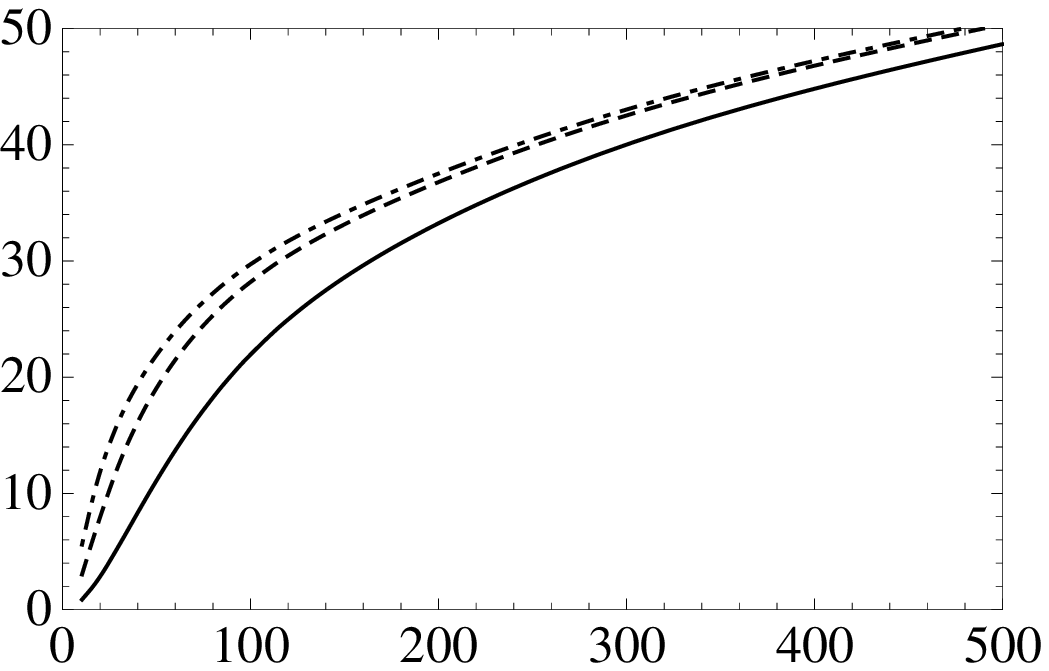}
\par\end{centering}

\begin{centering}
\begin{picture}(0,0)(0,0) \put(-20,10){ $T_{\mathrm{lab}}\,(\mathrm{MeV})$}
\put(-120,80){\rotatebox{90}{$t_{0}\,(\mathrm{hour})$}} \end{picture}
\par\end{centering}

\caption{\label{fig:t0}The dependence of $t_{0}$~(hour) on $T_{\mathrm{lab}}$~(MeV)
for $n=\unit[10^{14}]{cm^{-2}}$ and $f=\unit[10^{6}]{c^{-1}}$. The
acceptance angles in the lab frame are \mbox{$\theta_{\mathrm{acc}}^{(l)}=\unit[10]{mrad}$}
(solid curve), \mbox{$\theta_{\mathrm{acc}}^{(l)}=\unit[20]{mrad}$}
(dashed curve), \mbox{$\theta_{\mathrm{acc}}^{(l)}=\unit[30]{mrad}$}
(dashed\protect\nobreakdash-dotted curve).}

\end{figure}

Let us analyze the kinetics of polarization. Let $\boldsymbol{P}_{T}$
be the target polarization vector and $\boldsymbol{\zeta}_{T}=\boldsymbol{P}_{T}/P_{T}$.
Antiproton beam polarization vector $\boldsymbol{P}_{B}$ is collinear
to $\boldsymbol{\zeta}_{T}$ in both cases $\boldsymbol{\zeta}_{T}\perp\boldsymbol{v}$
and $\boldsymbol{\zeta}_{T}\parallel\boldsymbol{v}$. General solution
of the kinetic equation is given in Refs.~\cite{MilStr05,NikPav06}.
In our case, when only spin-\hspace{0cm}filtering mechanism is important,
we have

\begin{equation}
\begin{array}{l}
{\displaystyle P_{B}(t)=\tanh\left[\frac{t}{2}\left(\Omega_{-}^{out}-\Omega_{+}^{out}\right)\right],}\\
{\displaystyle N(t)=\frac{1}{2}N(0)\left[\exp\left(-\Omega_{+}^{out}t\right)+\exp\left(-\Omega_{-}^{out}t\right)\right],}\end{array}\label{eq:PN}\end{equation}
where

\begin{equation}
\Omega_{\pm}^{out}=nf\left\{ \sigma_{0}\pm P_{T}\left[\sigma_{1}+(\boldsymbol{\zeta}_{T}\cdot\boldsymbol{v})^{2}\left(\sigma_{2}-\sigma_{1}\right)\right]\right\} .\end{equation}
Here $n$ is the areal density of the target and $f$ is the beam
revolving frequency. It follows from our calculations that \mbox{$\left|\Omega_{-}^{out}-\Omega_{+}^{out}\right|\ll\left(\Omega_{-}^{out}+\Omega_{+}^{out}\right)$}
which allows us to simplify Eq.~\eqref{eq:PN}. The beam lifetime
due to the interaction with a target~is

\begin{equation}
\tau_{b}=\frac{2}{\Omega_{-}^{out}+\Omega_{+}^{out}}=\frac{1}{nf\sigma_{0}}.\end{equation}
Note that Figure Of Merit \mbox{$\mathrm{FOM}(t)=P_{B}^{2}(t)N(t)$}
is maximal at $t_{0}=2\tau_{b}$, when the number of antiprotons is
\mbox{$N(t_{0})\approx0{,}14N(0)$}. The polarization time $t_{0}$
as a function of the kinetic energy $T_{\mathrm{lab}}$ in the lab
frame is shown in Fig.~\ref{fig:t0}.

For the polarization degree at $t_{0}$, we have

\begin{equation}
P_{B}(t_{0})=\begin{cases}
-2P_{T}{\displaystyle \frac{\sigma_{1}}{\sigma_{0}}}, & \text{if}\:\boldsymbol{\zeta}_{T}\cdot\boldsymbol{v}=0,\\
-2P_{T}{\displaystyle \frac{\sigma_{2}}{\sigma_{0}}}, & \text{if}\,\left|\boldsymbol{\zeta}_{T}\cdot\boldsymbol{v}\right|=1.\end{cases}\end{equation}
Positive (negative) sign of $P_{B}(t_{0})$ means that the beam polarization
is parallel (antiparallel) to~$\boldsymbol{\zeta}_{T}$.

\begin{figure}
\begin{centering}
\hspace{2.8mm}\includegraphics[height=4.7cm]{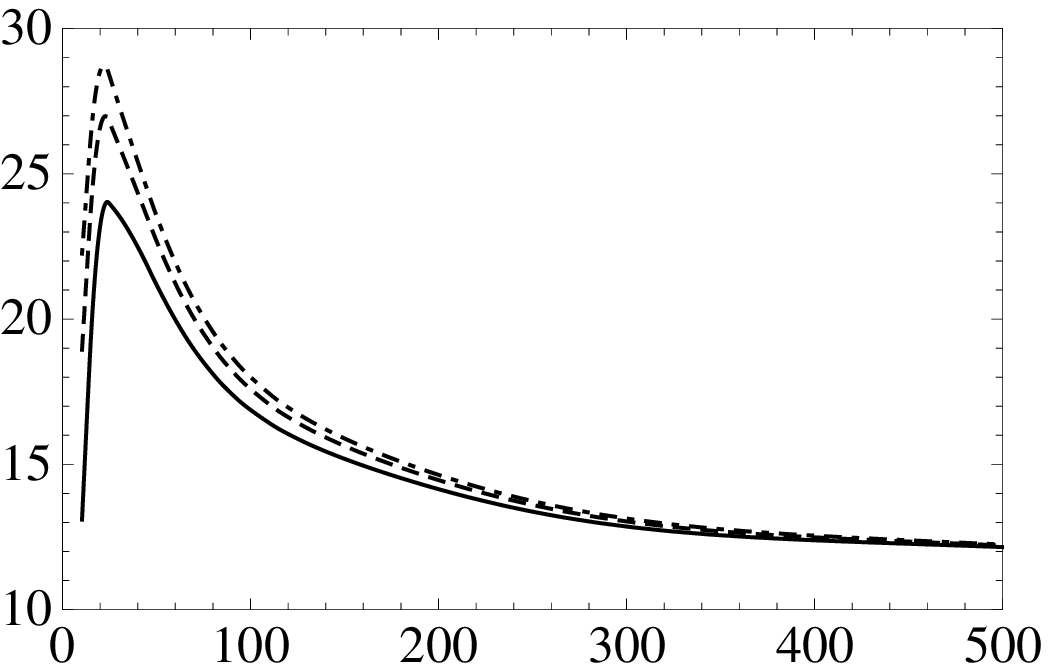}\hfill{}\includegraphics[height=4.7cm]{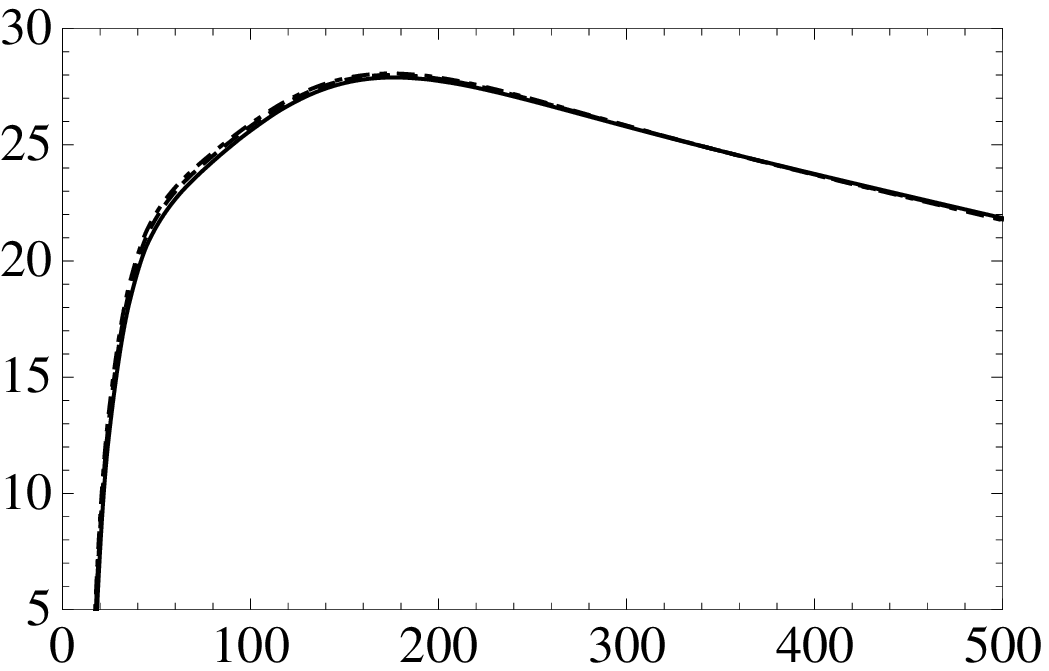}
\par\end{centering}

\begin{centering}
\begin{picture}(0,0)(0,0) \put(-145,10){ $T_{\mathrm{lab}}\,(\mathrm{MeV})$}
\put(110,10){ $T_{\mathrm{lab}}\,(\mathrm{MeV})$} \put(-250,80){\rotatebox{90}{$\sigma_{1}\,(\mathrm{mb})$}}
\put(10,80){\rotatebox{90}{$\sigma_{2}\,(\mathrm{mb})$}} \end{picture}
\par\end{centering}

\begin{centering}
\includegraphics[height=4.7cm]{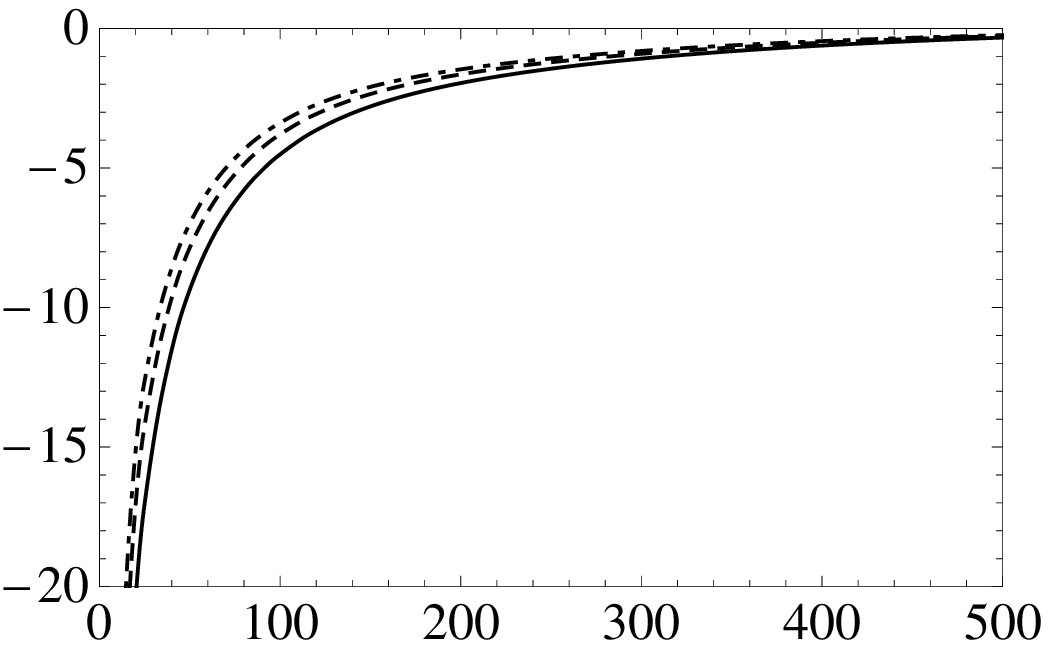}\hfill{}\includegraphics[height=4.7cm]{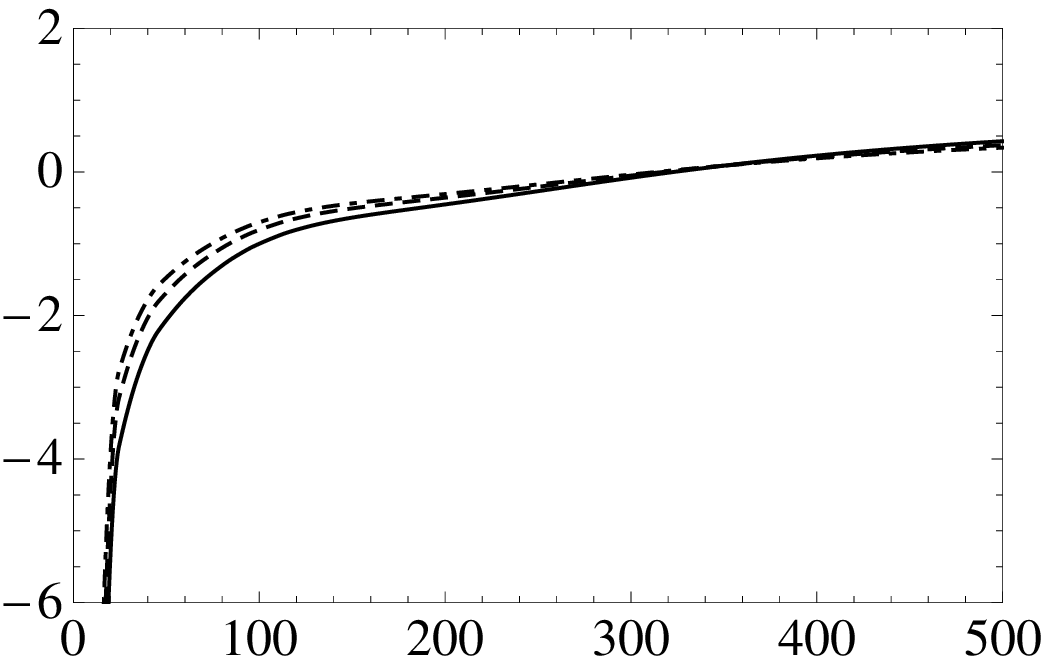}
\par\end{centering}

\begin{centering}
\begin{picture}(0,0)(0,0) \put(-145,10){ $T_{\mathrm{lab}}\,(\mathrm{MeV})$}
\put(110,10){ $T_{\mathrm{lab}}\,(\mathrm{MeV})$} \put(-250,75){\rotatebox{90}{$\sigma_{1}^{\mathrm{int}}\,(\mathrm{mb})$}}
\put(10,75){\rotatebox{90}{$\sigma_{2}^{\mathrm{int}}\,(\mathrm{mb})$}}
\end{picture}
\par\end{centering}

\caption{\label{fig:CrossSections}The dependence of $\sigma_{1}$, $\sigma_{2}$
and interference contributions $\sigma_{1}^{\mathrm{int}}$,~$\sigma_{2}^{\mathrm{int}}$~(mb)
on $T_{\mathrm{lab}}$~(MeV). The acceptance angles in the lab frame
are \mbox{$\theta_{\mathrm{acc}}^{(l)}=\unit[10]{mrad}$} (solid
curve), \mbox{$\theta_{\mathrm{acc}}^{(l)}=\unit[20]{mrad}$} (dashed
curve), \mbox{$\theta_{\mathrm{acc}}^{(l)}=\unit[30]{mrad}$} (dashed\protect\nobreakdash-dotted
curve).}

\end{figure}

The dependence of $\sigma_{1}$ and $\sigma_{2}$ on $T_{\mathrm{lab}}$
for different acceptance angles is shown in Fig.~\ref{fig:CrossSections}.
These cross sections depend on the acceptance angle completely due
the interference contribution. It was necessary to take interference
into account in the case of $pp$~scattering~\cite{MilStr05}. In
the case of $p\overline{p}$~scattering the interference contribution
is also important. Corresponding contributions $\sigma_{1,2}^{\mathrm{int}}$
are also shown in Fig.~\ref{fig:CrossSections}.

\begin{figure}
\begin{centering}
\includegraphics[height=4.7cm]{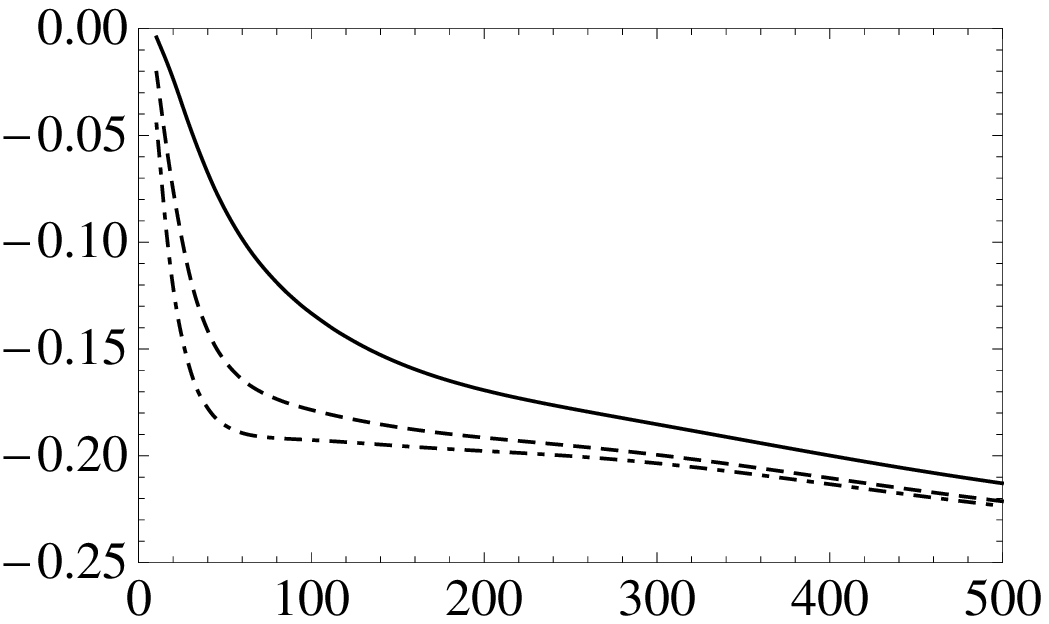}\hfill{}\includegraphics[height=4.7cm]{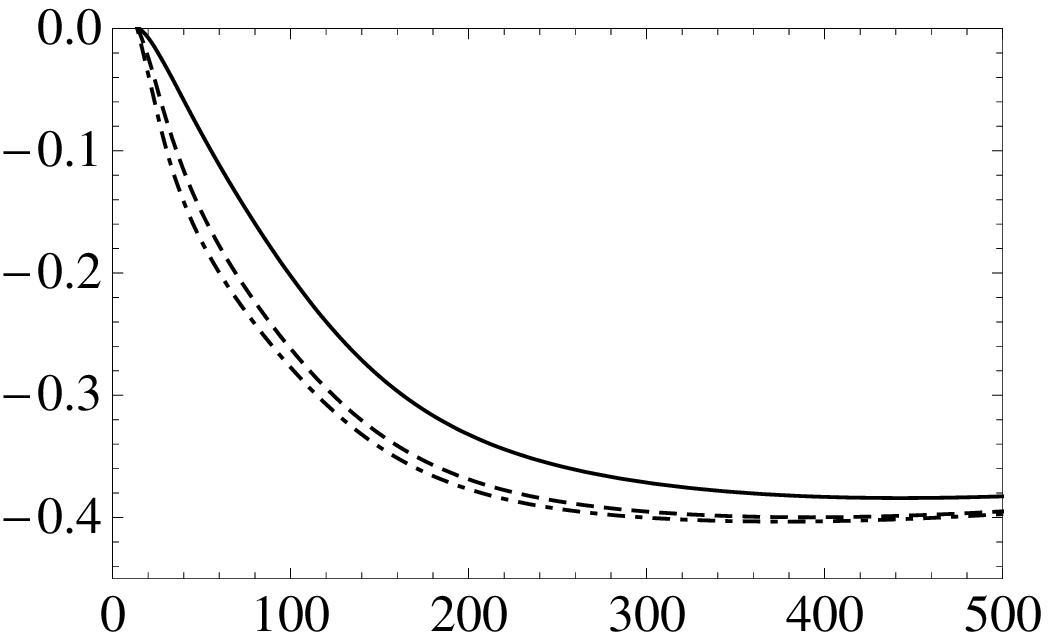}
\par\end{centering}

\begin{centering}
\begin{picture}(0,0)(0,0) \put(-140,10){ $T_{\mathrm{lab}}\,(\mathrm{MeV})$}
\put(110,10){ $T_{\mathrm{lab}}\,(\mathrm{MeV})$} \put(-250,85){\rotatebox{90}{$P_{\perp}$}}
\put(0,85){\rotatebox{90}{$P_{\parallel}$}} \end{picture}
\par\end{centering}

\caption{\label{fig:Polarizations}The dependence of $P_{B}(t_{0})$ for $P_{T}=1$
on $T_{\mathrm{lab}}$~(MeV) for \mbox{$\boldsymbol{\zeta}_{T}\cdot\boldsymbol{v}=0$
($P_{\perp}$)} and \mbox{$\left|\boldsymbol{\zeta}_{T}\cdot\boldsymbol{v}\right|=1$}
($P_{\|}$). The acceptance angles in the lab frame are \mbox{$\theta_{\mathrm{acc}}^{(l)}=\unit[10]{mrad}$}
(solid curve), \mbox{$\theta_{\mathrm{acc}}^{(l)}=\unit[20]{mrad}$}
(dashed curve), \mbox{$\theta_{\mathrm{acc}}^{(l)}=\unit[30]{mrad}$}
(dashed\protect\nobreakdash-dotted curve).}

\end{figure}

The dependence of polarization degree $P_{B}(t_{0})$ for $P_{T}=1$
on $T_{\mathrm{lab}}$~(MeV) for \mbox{$\boldsymbol{\zeta}_{T}\cdot\boldsymbol{v}=0$
($P_{\perp}$)} and \mbox{$\left|\boldsymbol{\zeta}_{T}\cdot\boldsymbol{v}\right|=1$}
($P_{\|}$) is shown in Fig.~\ref{fig:Polarizations}. In the case
$\boldsymbol{\zeta}_{T}\cdot\boldsymbol{v}=0$, the polarization degree
becomes independent of antiproton energy at $T_{\mathrm{lab}}$ about
$\unit[100]{MeV}$. With increasing acceptance angle the polarization
degree raises faster. In the case $\left|\boldsymbol{\zeta}_{T}\cdot\boldsymbol{v}\right|=1$,
the polarization degree increases slower but amounts to $40\%$ at
energy about $\unit[200]{MeV}$.

Let us compare our results with the previous calculations. In Ref.~\cite{DmitMilStr07},
spin-\hspace{0cm}dependent part of the cross section of $p\overline{p}$~scattering
has been calculated using Paris potential in energy range $\unit[20\div100]{MeV}$.
Authors predict positive $P_{\perp}$ with maximum $8\%$ at energy
about $\unit[60]{MeV}$ and negative $P_{\parallel}$, raising up
to $12\%$. Analogous calculations have been performed in Ref.~\cite{PAX09}
using several modifications of the Julich model. Note that the contribution
of interference between the Coulomb and strong amplitudes has been
omitted. Qualitatively, the dependence of polarization degree on the
antiproton energy, obtained in our paper, is similar to that in Ref.~\cite{PAX09},
but quantitative disagreement is rather big.

In conclusion, using the Nijmegen nucleon-\hspace{0cm}antinucleon
optical potential, we have calculated the spin-\hspace{0cm}dependent
part of the cross section of $p\overline{p}$~interaction and the
corresponding degree of the beam polarization. Our results indicate
that a fi{}ltering mechanism can provide a noticeable beam polarization
in a reasonable time. However, we state that today it is impossible
to predict the beam polarization with high accuracy because different
models give essentially different predictions. Only experimental investigation
of the spin-\hspace{0cm}dependent part of the cross section of $p\overline{p}$~scattering
can prove the applicability of potential models. Nevertheless, since
polarization degree in all models are rather big, we can hope that
filtering mechanism can be used to get polarized antiproton beam.
\begin{acknowledgments}
We are grateful to V. M. Strakhovenko for valuable discussions. The
work was supported in part by RFBR under Grant No.~09-02-00024.\end{acknowledgments}


\begin{thebibliography}{9}
\bibitem{PAX05}PAX Collaboration, \emph{Technical Proposal for Antiproton}-\hspace{0cm}\emph{Proton
Scattering Experiments with Polarization},~\href{http://arXiv.org/abs/hep-ex/0505054}{arXiv:hep-ex/0505054}~(2005).

\bibitem{MilStr05}A.~I.~Milstein, V.~M.~Strakhovenko, Phys.~Rev.~E\textbf{~72},~\href{http://pre.aps.org/abstract/PRE/v72/i6/e066503}{066503}~(2005).

\bibitem{NikPav06}N.~N.~Nikolaev, F.~F.~Pavlov, \emph{Polarization
Buildup of Stored Protons and Antiprotons: Filtex Result and Implications
for Pax at Fair, }\href{http://arxiv.org/abs/hep-ph/0601184}{arXiv:hep-ph/0601184}~(2006).

\bibitem{MilSalStr08}A.~I. Milstein, S.~G.~Salnikov, V.~M.~Strakhovenko,
Nucl.~Instr.~and Meth. in~Phys.~Res.~B~\textbf{266},~3453~(2008).

\bibitem{Csonka68}P.~L.~Csonka, NIM \textbf{63}, 247 (1968).

\bibitem{DmitMilStr07}V.~F.~Dmitriev, A.~I.~Milstein, V.~M.~Strakhovenko,
Nucl.~Instr.~and Meth. in~Phys.~Res.~B~\textbf{266},~1122~(2008).

\bibitem{PAX09}PAX Collaboration, \emph{Measurement of the Spin}-\hspace{0cm}\emph{Dependence
of the $\overline{p}p$~Interaction at the AD\nobreakdash-Ring},~\href{http://arxiv.org/abs/0904.2325v1}{arXiv:0904.2325}~(2009).

\bibitem{TimmRijSw94}R.~Timmermans, Th.~A.~Rijken, J.~J.~de~Swart,
Phys.~Rev.~C \textbf{50},~\href{http://prc.aps.org/abstract/PRC/v50/i1/p48_1}{48}~(1994).

\bibitem{NagRijSw78}M.~M.~Nagels, T.~A.~Rijken, J.~J.~de~Swart,
Phys.~Rev.~D \textbf{17},~\href{http://prd.aps.org/abstract/PRD/v17/i3/p768_1}{768}~(1978).
\end{thebibliography}
\end{document}